\documentclass[aps,prl,preprint,showpacs,amsmath,amssymb]{revtex4}

\usepackage[dvips]{epsfig}

\usepackage{natbib}
\bibpunct{(}{)}{,}{a}{}{,}

\newcommand{\pd}{{\phantom{\dagger}}}

\begin{document}

\title[]
{Dynamical Mean-Field Theory --- from Quantum Impurity Physics
to Lattice Problems}

\author{Ralf Bulla}
\affiliation{\mbox{Theoretische Physik III, Elektronische Korrelationen und
Magnetismus,}\\ \mbox{Universit\"at Augsburg,
86135 Augsburg, Germany}}

\begin{abstract}
Since the first investigation of the Hubbard model in the
limit of infinite dimensions by Metzner and Vollhardt, 
dynamical mean-field theory (DMFT) has become a very powerful
tool for the investigation of lattice models of correlated
electrons. In DMFT the lattice model is mapped on an effective
quantum impurity model in a bath which has to be determined
self-consistently. This approach lead to a significant progress 
in our understanding of typical correlation
problems such as the Mott transition; furthermore,
the combination of DMFT with ab-initio methods now allows
for a realistic treatment of correlated materials. 
The focus of these lecture notes is
on the relation between quantum impurity physics and the 
physics of lattice models within DMFT. Issues such as the
observability of impurity quantum phase transitions in the
corresponding lattice models are discussed in detail.

\end{abstract}


\maketitle

\section{Introduction}

Strongly correlated electron systems have been one of the most 
important topics in theoretical solid state research over the 
past decades. The major challenge in these systems is that interesting
features (like the Mott transition, \citet{imada,florian}) 
typically occur at intermediate
coupling strengths, so that perturbative approaches cannot be
applied. Although a number of {\em non}-perturbative techniques have
been developed for correlated electron systems, we are still far away
from a complete understanding of models like the two-dimensional
Hubbard model, let alone an exact solution of these models.

In the past 15 years, considerable progress has been achieved
through the development and application of the dynamical
mean-field theory (DMFT) \citep{MV,Georges}. 
One of the reasons for this success is that
in the standard implementation of DMFT, 
a lattice model is mapped on a quantum impurity model, for which
very powerful, non-perturbative methods are 
already available \citep{hew1}.
This quantum impurity model consists of a single correlated site
in a free fermionic bath whose structure has to be determined
self-consistently.

DMFT is exact in the limit of infinite coordination number
(or infinite dimensions). For any finite-dimensional system, DMFT
is an approximation as it fails to take into account non-local
fluctuations. Nevertheless, it can be used as a reasonable starting
point in many cases. In particular, a variety of 
cluster extensions of DMFT are 
currently developed
(for a review see \citet{qct}).

Extensive reviews on various aspects of DMFT have already been
published. The review by \citet{Georges} 
gives a detailed account
of the technical issues and early applications such as the
investigation of the Mott transition (see also the
lecture notes by \citet{Georges04} and more recent work
on the Mott transition \citep{GFC,BCV}).  A brief overview of DMFT
is given in
the Physics Today article by \citet{PhysicsToday}, which
also gives an introduction to the combination of ab-initio methods
with DMFT (for a review see \citet{ldaplusdmft}).

Here, we do not attempt to give an overview of all
the technical issues and the various applications of DMFT.
Instead, the focus of this paper is on the relation between the lattice
models and the quantum impurity models on which the lattice
model is mapped within DMFT.
Very often, this relation is seen from the following viewpoint:
let us start with a given lattice problem,
and within DMFT all that is left to do is to choose an
appropriate ``impurity solver'' 
which can then be used
as a black box within the DMFT self-consistency 
to arrive at the physics of the lattice model.
[It will be evident from reading the remainder of the paper that the
author regards the expression ``impurity solver'' as inadequate,
if not misleading, as it does neither give justice to the
technique one is refering to, which very often has its own
intellectual merits, nor to the effective impurity model
which deserves more attention instead of just being
``solved''.]
An alternative viewpoint is to start from quantum impurity models
in general, keeping in mind the enormous variety of physical
behaviour one can observe in these models, and then ask
the question whether this richness can also be observed in
the lattice counterparts within DMFT.

This is the viewpoint taken in this paper, discussed
in detail in the section entitled 
``DMFT and Quantum Impurity Physics''; 
there we introduce and classify a variety of
different quantum impurity models, with the impurity coupling
to fermionic and/or bosonic baths. Of particular importance
are impurity quantum phase transitions and their possible counterparts
in lattice models. Before that, we give a short introduction to DMFT
where we also discuss the emergence of a quantum
impurity model (the single-impurity Anderson model) in the
DMFT for the Hubbard model.
Finally, we summarize the main conclusions of the paper
and discuss the present status and future prospects of
DMFT.

\section{DMFT Essentials}

\begin{figure}[!t]
\epsfxsize=3.5in
\centerline{\epsffile{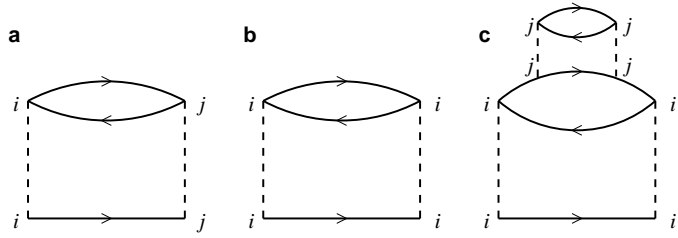}}
\caption{Self-energy diagrams for the Hubbard model.}

\label{fig:selfenergy}

\end{figure}

\citet{MV} have shown that 
a non-trivial limit of infinite spatial dimensions can be defined
for lattice
fermion models, provided that the model parameters are appropriately
scaled with the dimension. For the Hubbard model 
\citep{Hub,Gut,Kan}
\begin{equation}
   H = -t\sum_{<ij>\sigma} (c^\dagger_{i\sigma} c^\pd_{j\sigma} +
                   c^\dagger_{j\sigma} c^\pd_{i\sigma}) +
         U\sum_i c^\dagger_{i\uparrow} c^\pd_{i\uparrow}
            c^\dagger_{i\downarrow} c^\pd_{i\downarrow} \ ,
\label{eq:H}
\end{equation}
 this scaling only involves the hopping matrix
element $t=t^\ast/\sqrt{d}$, with $t^\ast$ fixed, while the local 
Coulomb repulsion
$U$ is unchanged. An important consequence of this scaling
is that the self-energy $\Sigma_{ij}(\omega)$ becomes
purely local. This can be understood from a simple counting argument
for the self-energy diagrams as shown, for example, 
in Fig.~\ref{fig:selfenergy}.

In this figure, the dashed lines denote the Coulomb repulsion $U$
and the solid lines the bare propagator $G(\omega)$. For $i$ and $j$
nearest neighbours in Fig. \ref{fig:selfenergy}a, the three propagators each contribute
a factor of $t$ to the diagram, i.e., a factor of $d^{-3/2}$.
With the number of nearest neighbours proportional to $d$, 
the sum of these self-energy
diagrams vanishes as $d^{-1/2}$ in the limit of $d\to\infty$.

Only local diagrams as in Fig. \ref{fig:selfenergy}b and c survive in the limit
$d\to\infty$. This does not mean that the Hubbard model is
reduced to a purely local model, as the propagation through
the lattice is still possible (dressed only with local self-energy terms,
as shown, for example, in Fig. \ref{fig:selfenergy}c).

The local self-energy of the Hubbard model obtained in this way
resembles that of a single-impurity Anderson model, which is local
anyway, as the Coulomb interaction $U$ in the Anderson model
only acts on a single site. The Hamiltonian of the 
single-impurity Anderson model reads:
\begin{eqnarray}
  H &=&   \sum_{\sigma} \varepsilon_{\rm f} f^\dagger_{\sigma}
                             f^\pd_{\sigma}
                 + U  f^\dagger_{\uparrow} f^\pd_{\uparrow}
                       f^\dagger_{\downarrow} f^\pd_{\downarrow}
                \nonumber \\
           & & + \sum_{k \sigma} \varepsilon_k c^\dagger_{k\sigma}
c^\pd_{k\sigma}
            +  \sum_{k \sigma} V
           \Big( f^\dagger_{\sigma} c^\pd_{k \sigma}
               +   c^\dagger_{k\sigma} f^\pd_{\sigma} \Big) \ . 
    \label{eq:siam}
\end{eqnarray}
With a proper definition of the bare propagator $\tilde{G}$
as
\begin{equation}
  \tilde{G}^{-1} = G_{\rm loc}^{-1} + \Sigma_{\rm HM} \ ,
\end{equation}
($G_{\rm loc}$ is the local Green function of the Hubbard model)
the self-energy functional $\Sigma_{\rm AM}[\tilde{G}]$ of
the Anderson model gives exactly the self-energy of the
Hubbard model.
The bare propagator $\tilde{G}$ itself depends on $\Sigma_{\rm HM}$
so that we arrive at a self-consistent procedure for the calculation of
the self-energy of the Hubbard model (for details, see \citet{Georges}).

Note that, although the above argument seems to be based
on a perturbative expansion of the self-energy, the
connection between the self-energies of the
impurity and the lattice model is {\em non}-perturbative.
This has been shown, for example, in \citet{MP} where
a non-perturbative construction of the Luttinger-Ward
functional has been used (another non-perturbative construction
is the cavity method discussed in \citet{Georges}). This means
that the mapping is valid in the whole parameter space
of the lattice model, even if the perturbative expansion
of the self-energy has a finite radius of convergence.

The non-trivial part of this self-consistent procedure is the calculation
of the self-energy of an effective single-impurity Anderson model,
for which the Green function for $U=0$ is equal to $\tilde{G}$.
This can be achieved using a variety of techniques
and the book of \citet{hew1} gives a fairly complete
overview of these methods in the context of the
Kondo problem. Note that the full frequency dependence
of the self-energy $\Sigma_{\rm AM}(\omega)$ needs to be calculated
for the effective impurity model.
Consequently, the Bethe ansatz \citep{hew1}
cannot be applied in the
context of the DMFT because its ability to calculate exact
results for quantum impurity problems is restricted to 
static quantities.

The following (incomplete) list includes those methods which have already
been applied within DMFT:
\begin{itemize}
  \item quantum Monte Carlo \citep{jarrell,Georges};
  \item iterated perturbation theory  \citep{GK,Georges};
  \item non-crossing approximation \citep{pru93,qct};
  \item exact diagonalization \citep{CF,Georges};
  \item numerical renormalization group \citep{BHP,BCV};
  \item local moment approach \citep{LMA1,LMA2};
  \item density matrix renormalization group 
         \citep{DDMRG1,DDMRG2};
  \item projective quantum Monte Carlo \citep{PQMC};
\end{itemize}
(for further information, please follow the references in brackets).
Interestingly,
the development of DMFT lead to a renewed search for methods to
calculate dynamic quantities of quantum impurity systems. New
developments are
the local moment approach,
the (dynamic) density matrix renormalization group, 
and the projector quantum Monte
Carlo method. For a recent
overview of techniques, in particular in the context of multi-band
models, see Sec.~III in \citet{qct}.

To conclude this section, we want to stress again the remarkable
fact that DMFT allows a mapping from a lattice model onto
a quantum impurity model; from this it follows immediately
that a thorough understanding of these quantum impurity models
is a prerequisite for a successful investigation of
lattice models within DMFT.

\section{DMFT and Quantum Impurity Physics}

\begin{figure}[!t]
\epsfxsize=2.7in
\centerline{\epsffile{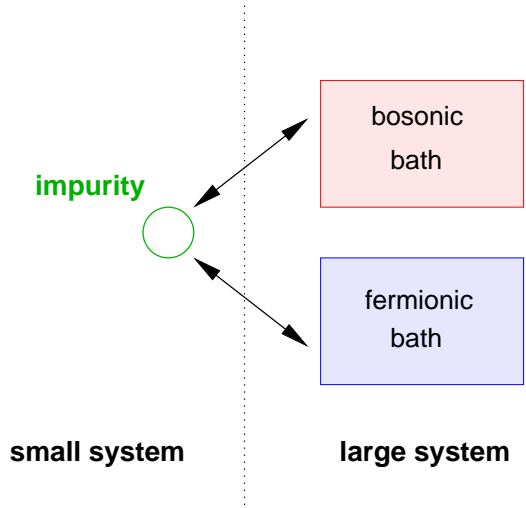}}
\caption{In a quantum impurity system a small system
(the impurity) is coupled to a large system of fermions and/or bosons
(the bath).}

\label{fig:qip}

\end{figure}

Consider a quantum system with a finite number of internal
degrees of freedom (the impurity) coupled to an infinite system
of non-interacting fermions or bosons with a continuous density
of states (the bath). An abstract view of such a 
{\em quantum impurity system} is given in Fig.~\ref{fig:qip}.
The impurity part of the Hamiltonian might have a complicated
structure (with, for example, two-particle terms due to the 
Coulomb repulsion between two fermions at the impurity site as
in eq.~(\ref{eq:siam})) but
we usually require the number of degrees of freedom of
the impurity to be small enough for a solution by exact
diagonalization. Due to the coupling between impurity and bath, 
the technical difficulty in solving
such a quantum impurity problem is given by both the
structure of the impurity term and the continuous spectrum
of the bath.

\begin{figure}[!t]
\epsfxsize=3.5in
\centerline{\epsffile{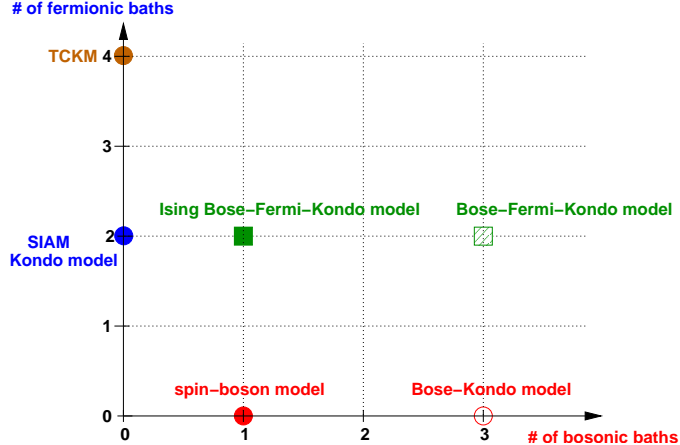}}
\caption{ Two-dimensional diagram for a classification
of quantum impurity systems by counting the number of fermionic
and bosonic baths to which the impurity couples.}

\label{fig:thediagram}

\end{figure}

A possible way to classify quantum impurity systems is to count
the number of baths to which the impurity couples. The bath
can be either fermionic or bosonic, so we arrive at a two-dimensional
diagram as in Fig.~\ref{fig:thediagram}.

The single-impurity Anderson model (SIAM) of eq.~(\ref{eq:siam})
occupies the point $(0,2)$ in this diagram as the impurity couples
to {\em two} fermionic baths (spin $\uparrow$ and
spin $\downarrow$), but no bosonic bath. As discussed above, the
single-impurity Anderson model 
appears as an effective model in the DMFT self-consistency
of the single-band Hubbard model. Multi-band generalizations
of the Hubbard model with, for example, an additional orbital
index $\alpha=1,\ldots,M$ 
map onto effective impurity models 
occupying the points  $(0,2M)$ in 
Fig.~\ref{fig:thediagram}.

One possible representative of the point $(1,0)$ in 
Fig.~\ref{fig:thediagram} is the spin-boson model
with the Hamiltonian:
\begin{equation}
H=-\frac{\Delta}{2}\sigma_{x}+\frac{\epsilon}{2}\sigma_{z}+
\sum_{i} \omega_{i}
     a_{i}^{\dagger} a^\pd_{i}
+\frac{\sigma_{z}}{2} \sum_{i}
    \lambda_{i}( a^\pd_{i} + a_{i}^{\dagger} ) \ .
\label{eq:sbm}
\end{equation}
This model naturally arises in the description of quantum
dissipative systems \citep{Leggett,Weiss}.
The dynamics of the two-state system, represented by the Pauli 
matrices $\sigma_{x,z}$,
is governed by the competition between the tunneling term $\Delta$ 
and the friction
term $\lambda_{i}(a^\pd_{i}+a_{i}^{\dagger})$;
the term $\frac{\epsilon}{2}\sigma_{z}$
represents an additional bias.
The operators $a_i^{(\dagger)}$ constitute a bath of harmonic oscillators responsible 
for the damping,
characterized by the bath spectral function
\begin{equation}
    J\left( \omega \right)=\pi \sum_{i}
\lambda_{i}^{2} \delta\left( \omega -\omega_{i} \right) \,.
\end{equation}
The so-called Bose-Kondo model \citep{Smith99,Sengupta} 
at $(3,0)$ can be viewed as a SU(2)-symmetric
generalization of the spin-boson model.

In the models at the points $(1,2)$ and $(3,2)$,  the impurity couples
to both fermionic and bosonic baths. These so-called Bose-Fermi Kondo
models appear in various physical systems and their connection to
DMFT is discussed further below.

It is important to note that each point in the diagram of 
Fig.~\ref{fig:thediagram} contains a variety of different models 
(with possibly different physical behaviour) as the diagram does
not distinguish between different structures of the impurity and different
densities of states of the bath. For example, the point
$(0,4)$ is occupied by the two-channel Kondo model (TCKM)  with
its well-known non-Fermi liquid fixed point \citep{CZ}, {\em and} the
SU(4) single-impurity Anderson model which has a Fermi liquid
ground state \citep{hew1}.

In any case, the classification of Fig.~\ref{fig:thediagram}
turns out to be useful for those (numerical) 
methods in which the computational effort depends
mainly on the number of baths (fermionic or bosonic). This holds
for those methods which are based on an actual diagonalization
of bath degrees of freedom (exact diagonalization,
numerical renormalization group, and
density matrix renormalization group). As an example, the
calculation of dynamic quantities within the
numerical renormalization group \citep{Wil75} is currently
restricted 
to the points $(1,0)$ \citep{BTV,BLTV}, 
$(0,2)$ \citep{Frota,Sakai,CHZ,BHP,hew1}, 
and $(0,4)$ \citep{two-band-nrg,and04}. 
On the other hand,
the number of baths does not play an essential role for methods
such as quantum Monte Carlo where all the bath degrees of freedom
are integrated out exactly and the calculations are performed
for an effective (although very complicated) action. In this case,
the number of degrees of freedom of the impurity site 
determines the computational effort (in addition to the values
of Coulomb repulsion and temperature).

One fascinating aspect of the models 
in Fig.~\ref{fig:thediagram} is the appearance of 
{\em impurity quantum phase transitions} due to the
competition of different physical mechanisms. A short
review on this topic is given
by \citet{BVhvar}.
Impurity quantum phase transitions are a special class of
so-called {\em boundary} quantum phase transitions 
(see Sec.~4 in \citet{MVreview})
where only the degrees of freedom of a subsystem
become critical 
(the impurity can be understood as a zero-dimensional boundary).
Such impurity quantum phase transitions require the
thermodynamic limit in the bath system, but are completely
independent of possible {\em bulk} phase transitions in the
bath.

Let us now come back to DMFT and the central issue of these
lecture notes: the relation between quantum impurity physics
and the physics of lattice models in DMFT. 
This issue shall be discussed in the context of the following
three questions:
\begin{enumerate}
  \item Is there a lattice counterpart for each
        possible type of quantum impurity model (which
        maps onto this quantum impurity model in DMFT)?
  \item Which of the interesting features of
        quantum impurity systems can be found
        within DMFT? In particular: which quantum
        phase transitions and quantum critical points
        ``survive'' the DMFT self-consistency?
  \item What are the new features of DMFT solutions
        for lattice models not present in the physics of
        quantum impurity models?
\end{enumerate}


Question No.~1 cannot be answered in general. The situation
is fairly clear for a class of models at
the points $(0,M)$. As briefly mentioned above, multi-band
Anderson models appear as effective impurity models
in the DMFT for multi-band Hubbard models.
Note that in this case the orbital structure
of the lattice model shows up in both the impurity part and in the
structure of the fermionic baths 
\citep{roz97,florens,ono,two-band-nrg}.

The two-channel
Kondo and Anderson lattices map on effective impurity models
at the point $(0,4)$ \citep{and97,and99}. The non-Fermi
liquid physics of the impurity models then shows up in
the corresponding lattice models as well.

Models at $(N,0)$ might appear as effective impurity models
in a DMFT treatment of the Bose-Hubbard model, a bosonic
version of the Hubbard model eq.~(\ref{eq:H}) with the fermionic operators
$c_{i\sigma}$ replaced by bosonic operators $b_{i\alpha}$ 
($\alpha=1,\ldots N$). This model has attracted renewed interest
in the context of ultra-cold (bosonic) atoms in optical
lattices \citep{Jaksch}. 
To our knowledge,
the Bose-Hubbard model has not yet been treated within DMFT.
The effective impurity models are presumably bosonic versions
of the single-impurity Anderson model, occupying the points
$(N,0)$ in the diagram of Fig.~\ref{fig:thediagram}.
In this context, the recently observed Mott transition
of ultra-cold atoms is of particular interest \citep{Greiner}; 
this Mott transition
should then correspond to an impurity quantum phase transition in
the bosonic single-impurity Anderson model;
one of the questions arising
here is whether a superfluid phase can at all be observed 
in such an effective impurity model. 

It is not clear whether a DMFT for these bosonic Hubbard models
would be as successful as the DMFT for fermionic models; but it
is definitely worth to go in this direction and thereby possibly widen
the range of applicability of the DMFT. Provided the basic
questions concerning the DMFT for the 
bosonic Hubbard model can be solved, one might even think of
an extension to mixtures of bosonic and fermionic atoms 
(see, for example, \citet{bfmixture} and references therein) which
would place
(within DMFT) the effective impurity models at the points
$(N,M)$.

Another route to effective impurity models with both fermionic and
bosonic baths is the {\em extended} DMFT \citep{eDMFT}. This specific
extension of DMFT is designed for the treatment of lattice
models with both local and non-local interaction terms.
A proper scaling of the non-local interaction term leads to
an effective impurity model where the impurity couples to
a fermionic bath (as usual) and, in addition, to a bosonic
bath corresponding to spin fluctuations of the surrounding
medium.

The extended DMFT has been applied to the Kondo lattice model
\citep{GS}, the periodic Anderson model \citep{sun}
(both supplemented by a coupling between spins on neighbouring
sites), the Hubbard model with long range Coulomb interactions
\citep{chitra},
and the $t$-$J$-model \citep{HRKW}.
In all these cases, the effective impurity model is
the Bose-Fermi Kondo model, with (depending on the details
of the problem) three bosonic baths (including the 
SU(2)-symmetric case), two bosonic baths (including the
XY-symmetric case) or one bosonic bath (the Ising case).
For the Ising case, the Bose-Fermi Kondo model takes the
form \citep{GS,ZD}:
\begin{eqnarray}
  H &=& J \vec{S}\cdot \vec{s} + 
          \sum_{k\sigma}\varepsilon_k c_{k\sigma}^{\dagger} c_{k\sigma}
\\ & +& 
\frac{\sigma_{z}}{2} \sum_{i}
    \lambda_{i}( a_{i} + a_{i}^{\dagger} ) +  \sum_{i} \omega_{i}
     a_{i}^{\dagger} a_{i} \ .
\end{eqnarray}
This model is characterized by the competition between screening
of the magnetic moment due to the coupling to the conduction
band, and the coupling to the bosons which favour an unscreened
spin. The resulting quantum phase transition and the connection
to local criticality will be discussed further below.

To summarize the discussion of question No.~1, we observe that
a full answer cannot be given at the moment and a lot of future
work is needed, in particular in the context of bosonic models.


From this discussion it is obvious that question No.~2 is
equally difficult to answer in general. 
Let us first consider the single-impurity Anderson model at $(0,2)$
with a general bath spectral function,
$\widetilde{\Delta}(\omega)=\pi V^2\sum_k \delta(\omega-\varepsilon_k)$,
in particular the case of a soft-gap at the Fermi level:
$\widetilde{\Delta}(\omega)=\Delta \vert\omega\vert^r$,
with an exponent $r>0$ \citep{withoff}.

\begin{figure}[!t]
\epsfxsize=2.9in
\centerline{\epsffile{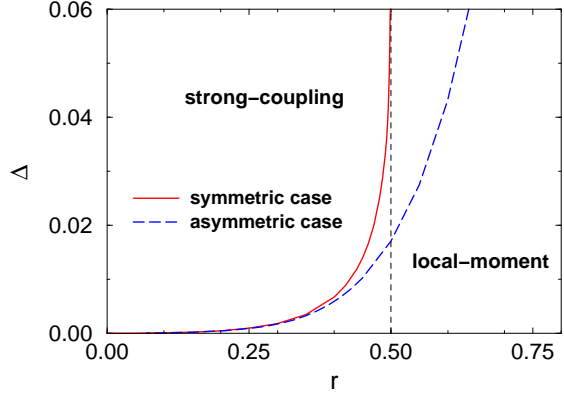}}
\caption{
      $T=0$ phase diagram for the soft-gap Anderson model
            in the p-h symmetric case (solid line, $U=10^{-3}$,
            $\varepsilon_f = -0.5 \cdot 10^{-3}$, conduction band
            cutoff at -1 and 1) and the p-h asymmetric case
            (dashed line,  $\varepsilon_f = -0.4 \cdot 10^{-3}$);
            $\Delta$ measures the hybridization strength,
            $\widetilde{\Delta}(\omega) =  \Delta |\omega|^r$
            \citep{GBI,bullapg}.
\vspace*{-0.3cm}
}
\label{fig:pd}
\end{figure}

The soft-gap case $0<r<\infty$ leads to a very rich behaviour,
in particular to a continuous transition
between a local-moment (LM) and a strong-coupling (SC) phase.
Figure \ref{fig:pd} shows a typical phase diagram for the soft-gap
Anderson model.
In the particle-hole (p-h) symmetric case 
(solid line) the critical coupling $\Delta_{\rm c}$
diverges at $r=\frac{1}{2}$, and no screening occurs for
$r>\frac{1}{2}$ (\citet{GBI,bullapg}).
No divergence occurs for p-h asymmetry (dashed line) \citep{GBI}.

For further details on the quantum critical properties and
the physics in the SC and LM phases see  
\citet{GLsoft,GBI,bullapg,MVreview,MVRB,MVLF}.

Important for the present discussion is the appearance
of a line of quantum critical points (solid and dashed lines in
Fig.~\ref{fig:pd}). Each quantum critical point on these
lines has a structure of excitations which differs from
both the SC and LM fixed points.

Obviously, the soft-gap models are only a sub-class of
single-impurity models with a non-constant density
of states. The question arises here whether such a
soft-gap density of states appears in the effective
impurity models within DMFT. So far, no example within
DMFT has been found in which a soft-gap stabilizes
itself under the DMFT iteration. Consequently, a
lattice analogue of the soft-gap critical points has not yet
been observed. (A soft-gap density of states might be
generated in the DMFT solution of the Hubbard-Holstein
model at the phase boundary between the metallic
and the bipolaronic insulating state \citep{hhm}.)

Neglecting for a moment the appearance of a line of quantum critical
points, the presence of both SC and LM phases in the impurity Anderson model
does have a counterpart in DMFT, which is the well studied
Mott transition in the Hubbard model. 

\begin{figure}[!t]
\epsfxsize=2.9in
\centerline{\epsffile{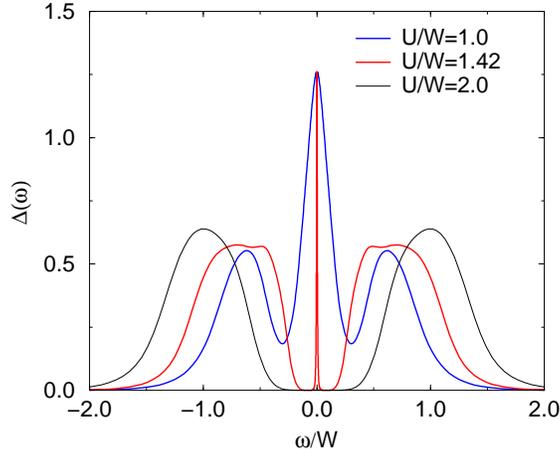}}
\caption{Spectral functions for the Hubbard model on a
  Bethe lattice for various values
  of $U$ \citep{rbmit}. A narrow quasiparticle peak develops at the
  Fermi level which vanishes at the critical $U_{\rm c}\approx 1.47W$
  (with $W$ the bandwidth of the non-interacting density
  of states).
}
\label{fig:mott}
\end{figure}

Figure \ref{fig:mott} shows the local spectral function for the
particle-hole symmetric Hubbard model at $T=0$ for
different values of the local Coulomb repulsion $U$.
These results are for the Bethe lattice (see Fig.~2 in 
\citet{rbmit});
in this case, the local spectral function is proportional
to the density of states of the free conduction band
appearing in the effective single-impurity Anderson model.

The Mott transition at $T=0$ is characterized by
a vanishing of the quasiparticle peak at a value of 
$U\approx 1.47W$. On the metallic side of the transition,
the $\Delta(\omega)$ approaches a constant for $\omega\to 0$,
in other words, the low frequency behaviour corresponds
to the $r=0$ limit of the soft-gap Anderson model where
no transition is observed and the system is always in the
SC phase. On the insulating side of the transition,
the $\Delta(\omega)$ shows a hard gap, corresponding to
the $r\to\infty$ limit of the soft-gap Anderson model.
For the p-h symmetric case, the system is always in the
LM phase for $r>1/2$ so we conclude that the insulating
side corresponds to the LM phase of the effective impurity
model. 

The Mott transition in DMFT shows the special feature that
the quasiparticle peak vanishes continuously whereas the
gap already has a finite value at the transition. 
Consequently, there is no quantum critical point
separating the metallic from the insulating side. Instead,
the transition point itself belongs to the insulating side
and is not characterized by a critical spectrum of
excitations.

The latter observation holds quite generally for all quantum
phase transitions observed in lattice models within
(standard) DMFT: no quantum critical behaviour has been
reported so far. In any case, critical behaviour in DMFT
cannot be due to non-local fluctuations; the only possibility
is {\em locally} critical behaviour which has indeed
been observed in the extended DMFT.

As discussed briefly above, a lattice model within extended
DMFT \citep{eDMFT} is mapped onto one of the variants of the
Bose-Fermi Kondo model. This model shows lines of
quantum critical points when the bath spectral functions
of the fermionic or bosonic baths follow a power-law
at low energies \citep{KV}.
The case of a sub-Ohmic
bosonic bath is of special importance here. Let us concentrate
on the Bose-Fermi Kondo model with Ising coupling to the
bosonic bath and a constant density of states of the conduction
electrons. This model can be mapped onto the spin-boson
model with a bath spectral function which is the sum
of an Ohmic (coming from the fermionic bath) and a sub-Ohmic
part \citep{ZGS}.
The quantum critical properties of this model are
the same as for the pure sub-Ohmic spin-boson model which
have been recently discussed in \citet{BTV} and \citet{VTB}. 
Similar to the 
soft-gap Anderson model discussed above, the sub-Ohmic 
spin-boson model shows a line of quantum critical points
for bath exponents $0<s<1$. The hyperscaling properties
of this model, and the corresponding $\omega/T$-scaling
over the whole range of $s$ values \citep{VTB}
have direct consequences
for the Kondo lattice model with Ising anisotropy.
As discussed in \citet{ZGS}, the criticality of the 
Bose-Fermi Kondo model is embedded in the criticality
of the magnetic quantum phase transition of the 
Kondo lattice model.


Let us now turn to question No.~3. Some of the features
of lattice models within DMFT {\em not} present in the 
physics of quantum impurity models are trivially connected
to the existence of a lattice. This allows, for example,
to define transport properties on the lattice
\citep{Georges,pjf}
and, in particular, the existence of long-range ordered
phases which cannot have a direct 
counterpart in impurity models.

A typical example are antiferromagnetic phases in the
Hubbard model which have been discussed in detail in
\citet{Georges,zpb}.
Such a symmetry breaking
does not exist in quantum impurity models. Furthermore,
phase transitions in quantum impurity systems are restricted
to {\em zero} temperature which is not the case in lattice
models within DMFT. It turns out, however, that the 
mean-field nature of the DMFT determines the behaviour
close to continuous (symmetry-breaking) transitions
both at zero and finite temperature. As discussed in detail
in \citet{byczuk}, only mean-field exponents for the vanishing of the
order parameter or the divergence of the susceptibility
can be observed, in agreement with numerical results 
\citep{jarrell,jp}.

Another example of a phase transition at finite temperature
is the first order Mott transition in the Hubbard model 
\citep{BCV,Georges,GFC}.
The line of first order transitions terminates at a 
finite-temperature critical point, which has been discussed in
detail in \citet{GFC}. Physical properties such as the temperature
dependence of the resistivity, $\rho(T)$, close to this
critical end-point cannot have a direct counterpart in
an effective impurity model. The reason is that
for each temperature, the DMFT self-consistency results
in a {\em different} quantum impurity model. The converged
effective impurity models therefore change with temperature
(and with all the model parameters).

\section{Conclusions}

The main focus of these lecture notes is on the relationship
between quantum impurity models and lattice models within
DMFT. Within the standard implementation of DMFT, a lattice
model is mapped on an effective impurity model supplemented
by a self-consistency condition. In this context, we discussed
questions such as the observability of impurity quantum
phase transitions in their lattice counterparts.

The DMFT for the simplest case where a lattice model maps
on the single-impurity Anderson model is very well developed.
Current topics of research are models involving additional
degrees of freedom, such as multi-orbital Hubbard models
in particular in the context of the LDA+DMFT approach
\citep{ldaplusdmft,JPSJ}, or
bosonic degrees of freedom in the Holstein-Hubbard model 
\citep{hhm}.
Another line of active research are extensions of the ``standard''
DMFT: cluster extensions as reviewed in \citet{qct}, or the so-called
extended DMFT \citep{eDMFT}. 

It should be mentioned here that other approaches exist
within DMFT which do not depend on the mapping
on an impurity model. One such approach is the Random
Dispersion Approximation \citep{rda} which has been
applied to the Mott transition in the Hubbard model;
in this approach the lattice model is simulated using
a cluster with random hoppings between all lattice sites.
Another more general attempt is based on the
Luttinger-Ward functional \citep{MP2}. Here the solution
of the lattice model is obtained via minimizing the
grand potential in a space of trial self-energies.

Despite these possible alternative approaches, further
progress in our understanding of quantum impurity systems
is certainly very important for the future development
of DMFT. This includes the
need to further improve existing techniques and 
possibly to invent new techniques to 
investigate quantum impurity systems, in particular
for systems with orbital and bosonic degrees of freedom.

Future work is necessary for a better understanding of
local criticality (within extended DMFT), an example of quantum critical
behaviour which, as discussed above, is found in both the
lattice model and the corresponding impurity model.
Some concerns about the ability of the extended DMFT to
explain lattice quantum critical points have been raised
recently \citep{KV}; furthermore, non-analyticities in the theory
might appear at low temperatures, as argued in \citet{HRKW}.

Future developments will certainly involve a fruitful
interplay between investigations of quantum impurity
models and lattice models within DMFT, and a variety of
new applications of DMFT, for example to bosonic systems,
are still lying ahead.

\section{Acknowledgments}

We thank
Krzysztof Byczuk,
Alex Hewson,
Marcus Kollar,
Thomas Pruschke,
Dieter Vollhardt,
and Matthias Vojta
for helpful discussions, and, in particular,
Zsolt Gulacsi for organization
of the 3rd Graduate School on Strongly Correlated Systems
in Debrecen.

This research was supported by the DFG through SFB 484.

\end{document}